\documentclass[aps,prl,twocolumn,groupedaddress,showpacs]{revtex4-2}

\usepackage[utf8]{inputenc}
\usepackage{amssymb}
\usepackage{amsmath}
\usepackage{mathrsfs}
\usepackage{color}
\usepackage{graphics, graphicx}
\usepackage{braket}

\usepackage[pdftitle={},
            pdfsubject={},
            pdfauthor={Steffen Backes},
            ]{hyperref}

\begin{document}

\author{Steffen Backes$^{1,2,3}$}
\email[]{steffen.backes@polytechnique.edu}
\author{Jae-Hoon Sim$^{1}$}
\author{Silke Biermann$^{1,2,3,4}$}
\affiliation{$^1$CPHT, CNRS, Ecole Polytechnique, Institut Polytechnique de Paris, Route de Saclay, 91128 Palaiseau, France}
\affiliation{$^2$Coll\`ege de France, 11 place Marcelin Berthelot, 75005 Paris, France}
\affiliation{$^3$European Theoretical Spectroscopy Facility, 91128 Palaiseau, France, Europe}
\affiliation{$^4$Department of Physics, Division of Mathematical Physics, Lund University, Professorsgatan 1, 22363 Lund, Sweden}

\date{\today}
\pacs{}

\title{Non-local Correlation Effects in Fermionic Many-Body Systems: Overcoming the Non-causality Problem}

\begin{abstract}
Motivated by the intriguing physics of quasi-2d fermionic systems, such as 
high-temperature superconducting oxides, layered transition metal chalcogenides 
or surface or interface systems, the development of many-body computational 
methods geared at including both local and non-local electronic correlations has 
become a rapidly evolving field. It has been realized, however, that the success 
of such methods can be hampered by the emergence of noncausal features in the 
effective or observable quantities involved. Here, we present a new approach of 
extending local many-body techniques such as dynamical mean field theory (DMFT) 
to nonlocal correlations, which preserves causality and has a physically 
intuitive interpretation. Our strategy has implications for the general class of 
DMFT-inspired many-body methods, and can be adapted to cluster, dual boson or 
dual fermion techniques with minimal effort. 
\end{abstract}

\maketitle

Electronic correlations arising from the Coulomb repulsion between electrons in 
a solid are at the heart of the astounding variety of emergent phenomena in 
condensed matter, ranging from exotic transport phenomena such as 
superconductivity to unconventional ordering phenomena involving spin, charge or 
orbital degrees of freedom.
Even in simplified lattice models with purely local interactions, the Coulomb
term requires refined approximations, such as 
dynamical mean-field theory (DMFT)\cite{Georges1992, GeorgesKotliar96,Vollhardt2010}.
This method proved to be highly successful in describing various effects of 
electronic correlations like the metal-insulator transition in transition metal 
compounds, Kondo physics, magnetic properties or 
superconductivity\cite{GeorgesKotliar96,Lichtenstein2000, 
Zolfl2001,Kotliar2006,ClusterDMFT,Aichhorn2009,Bauer2009,Medici2009,Haule2009, 
Hansmann2010,Yin2011, BiermannSpringer2014,Kitatani2015}.

Still, as a local approximation DMFT is inadequate for systems where nonlocal 
correlations and interactions are important. Thus, various methods aiming to 
reintroduce an approximate momentum dependence of the self-energy have been 
developed. Cluster 
extensions\cite{GeorgesKotliar96,Lichtenstein2000,ClusterDMFT} or the dynamical 
cluster approximation (DCA)\cite{DCA} introduce short range correlations, the 
dynamical vertex approximation (D$\Gamma$A)\cite{Toschi2007,Galler2017,scDGA}, 
TRILEX\cite{Ayral2015} or QUADRILEX\cite{Ayral2016b} schemes incorporate local 
but dynamical irreducible vertices, while dual fermion\cite{Rubtsov2008} or 
boson\cite{Lichtenstein2012} approaches perform a diagrammatic expansion around 
the DMFT solution to reintroduce momentum dependent correlation effects. 
Extended DMFT (EDMFT)\cite{Kotliar2002} incorporates screening effects of 
nonlocal interactions, while the combined GW+EDMFT 
method\cite{Biermann2003,Tomczak2012,Ayral2012,Ayral2013,Hansmann2013,Li2014, 
Biermann2014, Tomczak2014, Ayral2017} includes nonlocal screening and 
correlations on the level of the random phase and GW approximation. In this 
general approach one approximates the full self-energy by a sum of the local 
contribution $\Sigma_{loc}(\omega)$ generated from an effective impurity model, 
and a nonlocal part $\Sigma_{nonloc}(k,\omega)$, which for example can be 
obtained from a perturbative approach like the $GW$ approximation 
(GW+DMFT)\cite{Biermann2003,Biermann2014,Nilsson2017}, the fluctuation-exchange 
approximation (FLEX+DMFT)\cite{Aoki2015,Kitatani2015}, second order diagrams 
($\Sigma^{(2)}$+DMFT)\cite{Kananenka2015,Werner2015} or exact diagonalization 
(ED+DMFT)\cite{Liebsch2011}. A systematic formulation for such combination of 
local and nonlocal self-energies is provided by the
self-energy embedding theory (SEET)\cite{Kananenka2015,Zgid2017}.
While these methods proved highly successful in 
describing nonlocal phenomena like pseudo-gap physics, magnetism, 
superconductivity and charge 
order\cite{Zhang2007,Staar2013,Ayral2016,Stepanov2016,Sponza2017,Vucicevic2017, 
Lenz_2019,Li2020}, it has recently been realized that the combination of local 
techniques with nonlocal corrections can lead to noncausal, i.e. unphysical 
results, hampering the predictive power and applicability of the methods.

In this Letter we present a new self-consistent scheme for including nonlocal 
correlation effects into local theories, which preserves causality and has a 
transparent physical interpretation. In the limit of a local self-energy the 
standard DMFT equations are recovered. While we focus on GW+DMFT-like methods, 
this approach readily applies to cluster methods, and in general to all methods 
that include a feedback of the nonlocal on the local self-energy, like 
self-consistent dual boson, fermion or D$\Gamma$A\cite{scDGA} techniques. We 
benchmark our technique on an exactly solvable model, a two-site Hubbard dimer.

Noncausal behavior of physical quantities such as the Green's function, 
self-energy, hybridization or effective Weiss mean field corresponds to negative 
spectral weight in the spectral representation of their diagonal elements. In 
extensions of the dynamical cluster approximation(DCA) where the self-energy is 
interpolated to a continuous momentum-dependence, violation of causality has 
been reported\cite{Schulthess2020}. A failure of causality in GW+DMFT was 
reported for the screened Coulomb interaction in \cite{Chauvin2017} and in a 
two-atom system in the strongly correlated regime~\cite{Haule2017}. Here, the 
effective hybridization of the impurity model became significantly noncausal. For 
real materials Refs.\cite{Boehnke2016,Aryasetiawan2017,Werner2019} reported a 
general noncausal effective hybridization and interaction in GW+EDMFT. 
While non-causality was not 
detected in observable quantities \cite{Aryasetiawan2017}, inclusion of nonlocal 
correlations and screening appeared to lead to a decrease of local correlations 
in transition metal oxides compared to a local approximation. A spurious 
reduction of correlation strength due to a feedback of nonlocal correlations was 
also reported in self-consistent dual-fermion\cite{Ribic2018,Loon2018} and 
D$\Gamma$A\cite{scDGA} calculations, albeit causality violation was not 
investigated.

The problem of noncausality in these approaches is general and not related to 
the approximations involved. Indeed, it arises from imposing that the local part 
of the Green's function can be generated from a local model. This can be seen 
from the exact solution of a two-site Hubbard dimer with intersite hopping $t$ 
and onsite interaction $U$. Using $G_{imp}=G_{loc}$ the effective single-site 
impurity hybridization can be evaluated analytically as
\begin{align}
 \Delta(\omega) &= \frac{ (t-\Sigma_{inter}(\omega))^2}{\omega + \mu - \Sigma_{loc}(\omega)}, \label{eq:noncausal_dimer_hyb}
\end{align}
where $\Sigma_{inter}$ is the intersite and $\Sigma_{loc}$ the onsite 
self-energy. While this constitutes a causal hybridization in the DMFT limit 
$\Sigma_{inter}=0$, noncausal spectral weight $-\mathrm{Im}[\Delta]<0$ emerges 
in the general case including the exact solution, due to the imaginary term 
$\Sigma_{inter}$ in the nominator (see Fig.\ref{fig:general_dimer_hybridization} 
below). Thus, the noncausality problem due to the inclusion of a nonlocal 
self-energy is inherent to the commonly used form of the self-consistency 
equations, as even the exact solution generates a noncausal impurity model. 

For simplicity we consider a single-orbital fermionic Hubbard model with 
intersite hoppings $t_{ij}$, chemical potential $\mu$, and on-site interaction 
$U$. (The generalization to the multi-orbital case is straight forward.) The 
full action $S$ for this system is given by 
\begin{align}
S 
&= \int_0^{\beta} \mathrm{d}\tau \sum_{i\sigma} c^*_{i\sigma}(\tau)\left[ \frac{\partial}{\partial \tau} - \mu \right] c_{i\sigma}(\tau)   \\
    & -\sum_{ij,\sigma} \int_0^{\beta} \mathrm{d}\tau \ t_{ij} c^*_{i\sigma}(\tau)  c_{j\sigma}(\tau)  \nonumber 
+ \sum_{i} U \int_0^{\beta} \mathrm{d}\tau \ n_{i\uparrow}(\tau) n_{i\downarrow}(\tau),
\end{align}
where $c_{i\sigma}(\tau),c^*_{i\sigma}(\tau)$ are anticommuting Grassman 
variables, $n_{i\sigma}(\tau) = c^*_{i\sigma}(\tau)c_{i\sigma}(\tau)$ and 
$n_{i}(\tau) = n_{i\uparrow}(\tau) + n_{i\downarrow}(\tau)$. Following the idea 
of the cavity construction as outlined in Ref.\onlinecite{GeorgesKotliar96}, we 
split the action $S$ into three contributions $ S = S_0 + \Delta S + S^{(0)} $, 
where $S_0$ is the term that contains only the local quantities on site $i=0$, 
$\Delta S$ contains all contributions that couple the site $i=0$ to all other 
sites, and $S^{(0)}$ contains all the contributions of the lattice with site 
$i=0$ and its bonds connecting it being removed. Integrating out all the degrees 
of freedom except the ones on site $i=0$ one obtains an effective action of the 
form 
\begin{align}
 S_{eff} &= S_0
 - \sum_{n=1}^{\infty} \frac{(-1)^n}{n!} \braket{ (\Delta S)^n }^{(0)},
 \label{eq:eff_action}
\end{align}
where $\braket{  }^{(0)}$ indicates a trace over the system without the site 
$i=0$. In Refs.\onlinecite{MetznerVollhardt89, GeorgesKotliar96,Lleweilun1996} 
it has been shown that for infinite lattice connectivity and a dimensional 
rescaling of the hopping parameters only the term $n=2$ survives in the 
effective action. Therefore, in this limit the effective action reduces to the 
local part and the second order contribution, which takes the form of an 
effective impurity action parameterized by the effective Weiss field
\begin{align}
 \mathscr{G}^{-1}(\tau_1-\tau_2) 
 &= -\delta(\tau_1-\tau_2)\left[\frac{\partial}{\partial \tau} - \mu +t_{00} \right] \nonumber \\
 &\hspace*{1cm}- \sum_{ij\neq 0} t_{0i} t_{j0} G_{ij}^{(0)}(\tau_1-\tau_2) \label{eq:bath_from_cavity}.
\end{align}
Inserting the equality $G^{(0)}_{ij} = G_{ij} - G_{i0}G^{-1}_{00}G_{0j}$ into 
Eq.\eqref{eq:bath_from_cavity} and performing a Fourier transform, we obtain
\begin{align}
 \mathscr{G}^{-1} 
 &= i\omega + \mu - \braket{\epsilon} - \Delta(i\omega) \nonumber \\
 \mbox{with } 
 \Delta(i\omega) 
 &=   \braket{ \epsilon G \epsilon } - \braket{\epsilon G} \braket{G}^{-1}  \braket{ G \epsilon }  
\label{eq:eff_dyson_for_bath} ,
 \end{align}
 where the spin index $\sigma$ is suppressed for readability. The bracket 
$\braket{}=\int \mathrm{d}k$ indicates a local projection of the corresponding 
lattice quantity. $\epsilon(k)$ is the Fourier transform of the hoppings 
$t_{ij}$. We point out that Eq.\eqref{eq:bath_from_cavity} and 
Eq.\eqref{eq:eff_dyson_for_bath} are equivalent, even when the 
self-energy is nonlocal, and the only approximation has been performed on the 
effective action in Eq.\eqref{eq:eff_action}. Most importantly, since  
$G_{ij}^{(0)} $ is a causal Green's function by definition, the resulting Weiss 
field $\mathscr{G}$ and hybridization in this form are also causal. While in 
infinite dimensions the self-energy indeed does become local, for finite 
dimensions this constitutes an additional approximation, which simplifies 
Eq.\eqref{eq:eff_dyson_for_bath} to the local Dyson equation
$ \mathscr{G}^{-1}  = \braket{G}^{-1} + \braket{\Sigma} $
used in DMFT and its nonlocal extensions.

Now we explicitly consider a momentum dependent self-energy $\Sigma(k,i\omega)$ 
in the interacting Green's function 
 \begin{align}
G(k,i\omega) &= \left[ i\omega + \mu -\epsilon(k) - \Sigma(k,i\omega) \right]^{-1}.
\label{eq:G_lattice}
\end{align}
Rewriting Eq.\eqref{eq:eff_dyson_for_bath} without a local self-energy approximation
we arrive at a generalized cavity equation
\begin{widetext}
 \begin{align}
\mathscr{G}^{-1}
&= \braket{G}^{-1} + \braket{\Sigma} 
  - \Big( \braket{\Sigma G \Sigma} - \braket{\Sigma G} \braket{G}^{-1} \braket{G\Sigma}
     + 2\braket{\Sigma} - \braket{\Sigma G} \braket{G}^{-1} - \braket{G}^{-1}\braket{G\Sigma}\Big) .
     \label{eq:eff_dyson_for_bath_sigma}
\end{align}
\end{widetext}
Here one identifies the first two terms as the local Dyson equation, but with an 
additional correction term. As an exact rewriting of the causal equation 
\eqref{eq:eff_dyson_for_bath}, this expression is causal as well. Since the 
Dyson equation yields a noncausal bath in general as demonstrated in 
Eq\eqref{eq:noncausal_dimer_hyb}, the correction term is responsible for 
ensuring causality in the case of a nonlocal self-energy. This result can be 
extended to nonlocal screening and interactions, resulting in very similar 
equations for the effective impurity 
interaction~\cite{supplement,BackesNonlocalScreening}. Neglecting the 
momentum-dependence of the self-energy in the DMFT limit the local projection 
factorizes $\braket{G\Sigma}= \braket{G}\braket{\Sigma}$ and the correction 
terms vanishes, recovering the DMFT local Dyson equation. Even when the nonlocal 
self-energy is frequency-independent, this correction term does not vanish, e.g. 
in the case of a static Fock-like self-energy, but still modifies the bath.

GW+DMFT and related self-consistent schemes that include a nonlocal self-energy 
and enforce $G_{loc}=G_{imp}$, i.e. the impurity bath $\mathscr{G}(i\omega)$ is 
given by the local Dyson equation, will in general produce a noncausal solution, 
unless the correction term above is included. In cluster approaches such as cluster-DMFT, 
where the self-energy is local on a cluster and correlations within the cluster 
are included as offdiagonal elements, the local Dyson equation is causal. If the 
self-energy is periodized during the self-consistency to reestablish 
translational invariance, it picks up a true momentum dependence and 
noncausality emerges, unless the correction term is included. Similarly, the 
noncausality problem in the DCA for momentum-interpolated self-energies can be 
remedied by inclusion of the additional term above.
 
The correction term has a clear physical interpretation: redefining the hopping 
amplitudes such that they absorb the nonlocal self-energy $\tilde{t}_{ij} = 
t_{ij}  + \Sigma_{ij}, \ i\neq j$, one recovers the local Dyson equation, since 
the remaining self-energy is purely local. This implies that the Dyson equation 
dresses all intersite hoppings with the nonlocal self-energy, including the ones 
connecting the impurity with the bath $t_{i0},t_{0j}$. But the cavity 
construction and Eq.\eqref{eq:bath_from_cavity} require the bare hopping 
amplitudes for connecting the impurity with the bath. Therefore, the additional 
correction term in Eq.\eqref{eq:eff_dyson_for_bath_sigma} removes the nonlocal 
self-energy contribution dressing the bath-impurity hoppings, which would 
otherwise result in a 'double-counting' of the nonlocal self-energy effects
on the local correlations.

When deriving DMFT within a functional approach\cite{Potthoff2003,Kotliar2006}, 
the resulting stationarity condition $G_{loc}=G_{imp}$ is equivalent to the 
cavity construction in infinite dimensions, and thus the bath is always related 
to a physical lattice via a second order approximation of the effective action. 
When including a nonlocal self-energy as in the standard GW+DMFT 
scheme\cite{Biermann2014,Aryasetiawan2017} or related methods, this connection 
to a lattice via the cavity construction is sacrificed.
Here, via Eq.\eqref{eq:eff_dyson_for_bath_sigma} we restrict 
ourselves to the solutions that retain the relation to a physical lattice via 
the cavity construction, and thus preserve causality. This approach results in 
$G_{imp} \neq G_{loc}$, as can be seen by comparing 
Eq.\eqref{eq:eff_dyson_for_bath_sigma} with the impurity Dyson equation, a 
feature common to diagrammatic extensions of 
DMFT\cite{Gukelberger2017,Loon2016,Rohringer2016,Krien2017}. Now the impurity 
Green's function becomes an auxiliary, albeit always causal quantity that 
generates an approximation to the local self-energy. Assuming 
$\Sigma_{loc}=\Sigma_{imp}$, this establishes the self-consistency relation for 
the local self-energy  used to solve Eq.\eqref{eq:eff_dyson_for_bath_sigma} 
iteratively.
For a given $\Sigma_{nonloc}$ the resulting scheme has the following form:

1) Make a starting guess for $\Sigma_{loc}$.  

2)  Deduce $G(k,i\omega)$ from Eq.\eqref{eq:G_lattice} with $\Sigma=\Sigma_{loc}+\Sigma_{nonloc}$

3)  Obtain $\Delta(i\omega)$ and $\mathscr{G}(i\omega)$ from Eq.\eqref{eq:eff_dyson_for_bath} 

4)  Solve impurity model to obtain $G_{imp}$ and deduce \\
    \hspace*{0.7cm}   $\Sigma_{imp}= \mathscr{G}^{-1} - G^{-1}_{imp}$ 
       
5)  Iterate using $\Sigma_{loc}=\Sigma_{imp}$ until self-consistency.

\begin{figure}[t]
\includegraphics[width=0.5\textwidth]{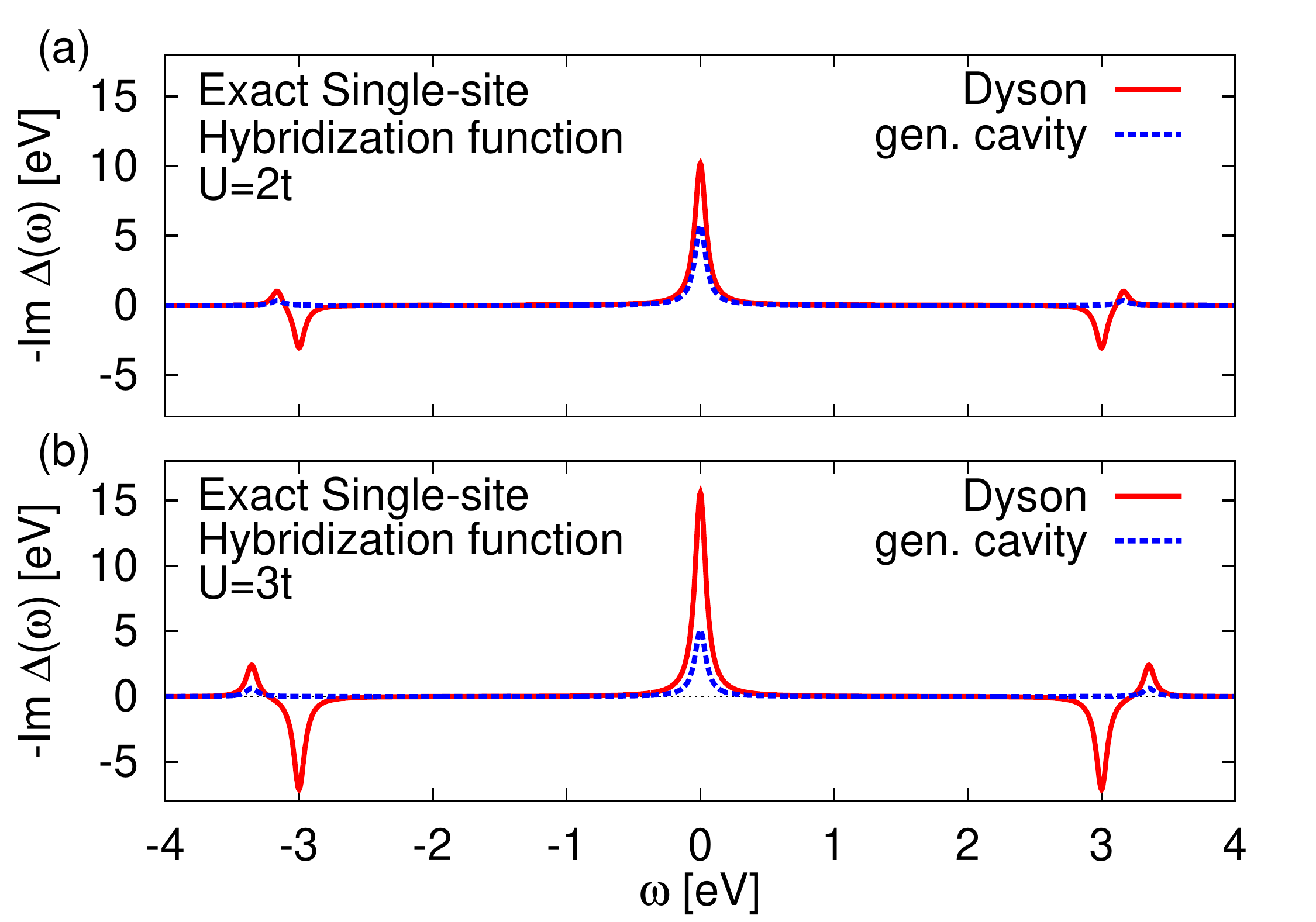} 
\caption{The effective single-site impurity hybridization for the exact solution 
of a two-site dimer at half filling for the interaction (a) $U=2t$ and (b) 
$U=3t$. Noncausal spectral weight is evident by negative values of 
$-\mathrm{Im}\Delta$ in the bath generated from Eq.\eqref{eq:noncausal_dimer_hyb}.
The generalized cavity equation Eq.\eqref{eq:causal_dimer_hyb} generates a bath 
that is always causal. The hybridization at $\omega=0$ is overestimated in the 
Dyson equation, leading to a reduced strength of electronic correlations.
}
\label{fig:general_dimer_hybridization}
\end{figure}

\noindent We now apply this scheme to the two-site dimer. Using 
Eq.\eqref{eq:eff_dyson_for_bath} or Eq.\eqref{eq:eff_dyson_for_bath_sigma} we 
analytically evaluate the effective single-site hybridization and obtain
\begin{align}
 \Delta(\omega) &= \frac{ t^2}{\omega + \mu - \Sigma_{loc}(\omega)}. \label{eq:causal_dimer_hyb}
\end{align}
This expression is always causal and the intersite hopping is not renormalized 
by the intersite self-energy, in contrast to the Dyson equation in 
Eq.\eqref{eq:noncausal_dimer_hyb}. This confirms our interpretation of the 
correction term in Eq.\eqref{eq:eff_dyson_for_bath_sigma}, which ensures that 
the impurity is coupled to the bath via the bare hopping $t$. As there is only 
one bond in the dimer, the bath consists only of the other site, thus no 
intersite self-energy enters the bath. The Dyson equation includes the intersite 
self-energy on the bond, which gives rise to the noncausal spectral weight, as 
its effect is similar to a dissipative term in the one-particle Hamiltonian.

In Fig.\ref{fig:general_dimer_hybridization} we show the effective single-site 
impurity hybridization, generated from Eqs.\eqref{eq:noncausal_dimer_hyb} and 
\eqref{eq:causal_dimer_hyb} for $U=2t$ and $3t$, using the exact self-energy. 
The Dyson equation generates noncausal spectral weight at $\omega=\pm 3$~eV, 
enhanced for larger interactions. The new scheme always produces a causal 
result. At the Fermi level $\omega=0$ the Dyson equation significantly 
overestimates the hybridization compared to Eq.\eqref{eq:causal_dimer_hyb}, 
which we also observed for other model systems and thus expect to be a general 
effect. This can explain the reported reduced strength of electronic 
correlations~\cite{Aryasetiawan2017,Ribic2018,Loon2018,Schulthess2020,scDGA}.

\begin{figure}[t]
\includegraphics[width=0.5\textwidth]{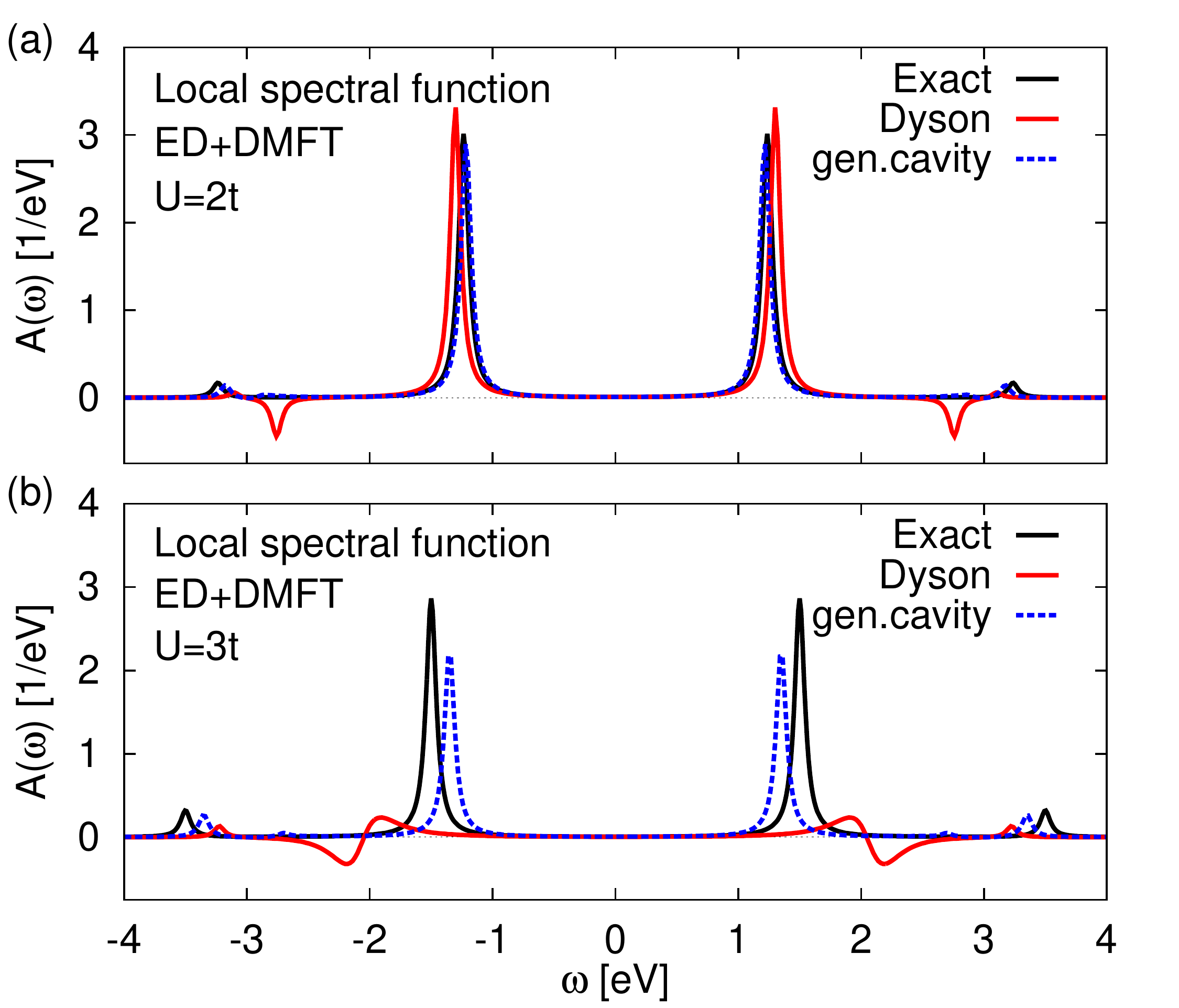} 
\caption{The ED+DMFT local spectral function for a two-site dimer at half 
filling, compared to the exact solution for (a) $U=2t$ and (b) $U=3t$. The exact 
form of the nonlocal self-energy has been used. The generalized cavity equations 
in Eq.\eqref{eq:eff_dyson_for_bath} reproduce the exact solution well. For small 
interaction values the Dyson equation generates noncausal spectral weight, and 
fails completely for $U \gtrsim 3t$, as the Green's function becomes nonanalytic 
(see explanation in main text).
}
\label{fig:general_spec_dimer}
\end{figure}
In Fig.\ref{fig:general_spec_dimer} we show converged results for an ED+DMFT 
scheme, using the exact nonlocal self-energy and determining the local 
self-energy self-consistently. We employed an exact diagonalization impurity 
solver allowing for a non-hermitian bath to describe the noncausal impurity 
hybridization. Both the generalized cavity scheme and Dyson equation agree well 
with the exact result for $U=2t$, with the latter showing a slightly worse 
agreement and noncausal spectral weight. For $U \gtrsim 3t$ the Dyson equation 
fails completely while the cavity scheme is still in qualitative agreement with 
the exact solution. The failure of the Dyson equation can be traced back to the 
poles of the Green's function separating from the real axis and entering the 
complex plane, corresponding to imaginary Eigenvalues that arise from a 
noncausal bath, i.e. nonhermitian hybridization amplitudes\cite{supplement}. In 
this case the Green's function is no longer analytic in either the upper/lower 
complex plane, violating a necessary condition for the Hilbert transform, which 
connects the spectra on the real to the imaginary Matsubara axis. As a result, 
analytic continuation such as the maximum entropy method\cite{Maxent} can no 
longer be applied\cite{supplement}.

\begin{figure}[t]
\includegraphics[width=0.5\textwidth]{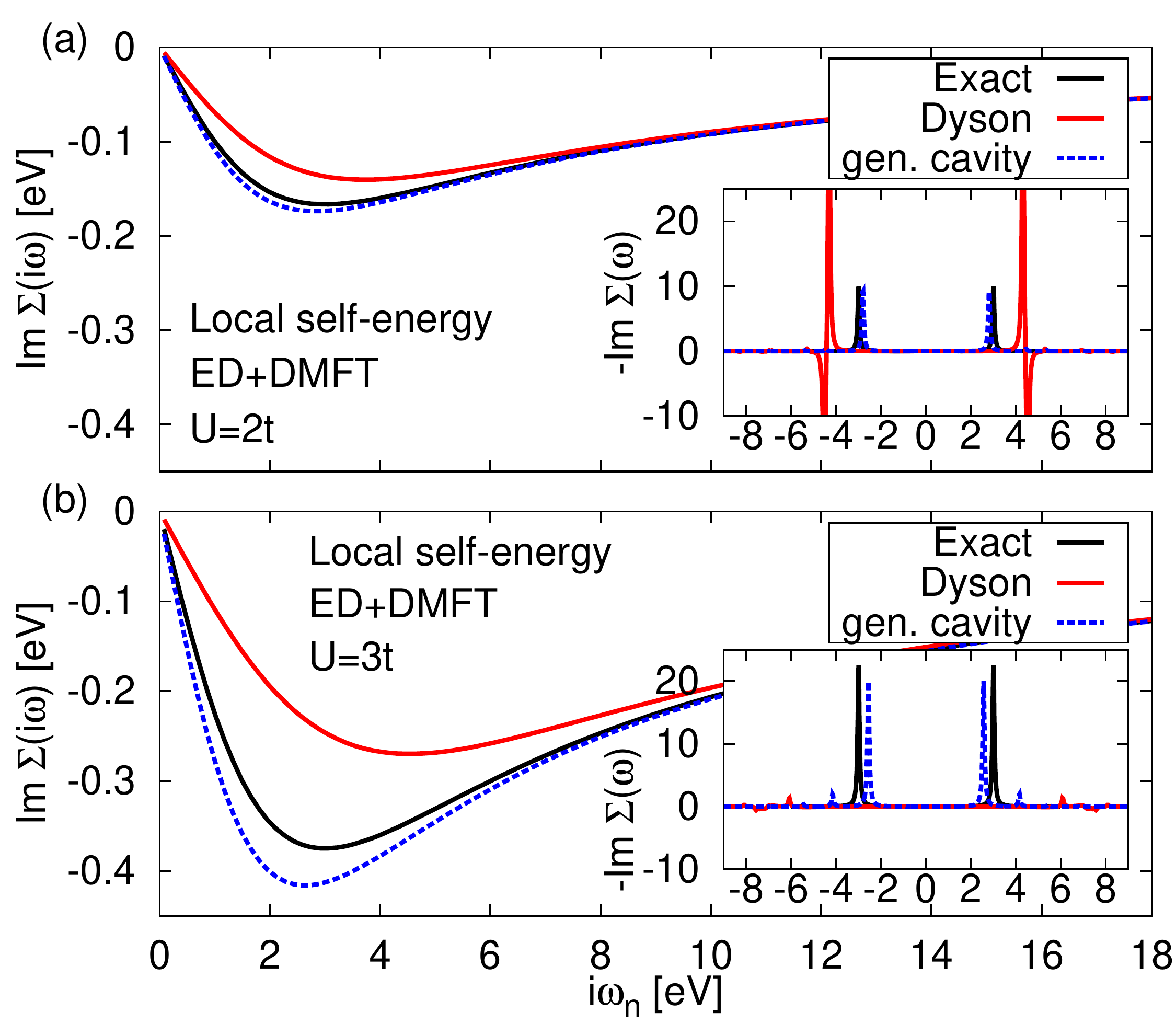} 
\caption{The local self-energy $\Sigma_{loc}$ on Matsubara $i\omega_n$ or real 
frequencies $\omega$ (inset) for the dimer as in 
Fig.\ref{fig:general_spec_dimer}. The generalized cavity equations obtain a 
self-energy in good agreement with the exact result. The local Dyson equation 
underestimates the correlation strength, creates a noncausal self-energy and 
breaks down for $U \gtrsim 3t$ (see explanation in main text).
}
\label{fig:general_sigma_dimer}
\end{figure}

This is evident in the local self-energy as well 
(Fig.\eqref{fig:general_sigma_dimer}). While for the Dyson equation 
$\Sigma_{loc}$ vanishes on the real axis for $U \gtrsim 3t$ as all poles have 
left the real axis, the Matsubara self-energy is finite and shows no obvious 
abnormal features apart from an underestimation of correlation strength. The 
lack of any signatures in the Matsubara data makes this especially critical for 
real materials calculations, where real-frequency impurity solvers are often not 
feasible, making it impossible to detect such a breakdown of the formalism. The 
generalized cavity equations instead generate a causal self-energy that agrees 
well with the exact solution.

We have presented a generalized form of a cavity construction for correlated 
fermionic systems that explicitly includes both local and nonlocal correlations. 
The result is an effective impurity problem that has a physically meaningful 
interpretation and solves the noncausality problem that so far has hampered the 
development of many-body methods for describing non-local correlations. The 
method can be readily adapted to any method which incorporates nonlocal 
self-energies in the form of an effective impurity problem, such as cluster 
methods, self-consistent dual fermion/boson or D$\Gamma$A techniques. Using a 
two-site dimer as a benchmark system, the scheme results in causal solutions in 
very good agreement with the exact result. This provides important progress as 
compared to current state-of-the-art approaches, which generate noncausal 
spectral weight, underestimate the correlation strength and violate analyticity. 
Our scheme can be readily extended to nonlocal screening and interactions, as 
discussed in the supplementary 
material~\cite{supplement,BackesNonlocalScreening}.

\acknowledgments 

The authors gratefully acknowledge discussions with 
Hartmut Hafermann, 
Hong Jiang, 
Aaram Kim, 
Benjamin Lenz, 
and Lucia Reining. This work has been supported 
by a Consolidator Grant of the European Research Council (Project 
CorrelMat-617196) and IDRIS/GENCI under project number t2020091393.
We are grateful to the CPHT computer support team.



%

 \end{document}


\author{Steffen Backes$^{1,2,3}$}
\email[]{steffen.backes@polytechnique.edu}
\author{Jae-Hoon Sim$^{1}$}
\author{Silke Biermann$^{1,2,3,4}$}
\affiliation{$^1$CPHT, CNRS, Ecole Polytechnique, Institut Polytechnique de Paris, Route de Saclay, 91128 Palaiseau, France}
\affiliation{$^2$Coll\`ege de France, 11 place Marcelin Berthelot, 75005 Paris, France}
\affiliation{$^3$European Theoretical Spectroscopy Facility, 91128 Palaiseau, France, Europe}
\affiliation{$^4$Department of Physics, Division of Mathematical Physics, Lund University, Professorsgatan 1, 22363 Lund, Sweden}

\date{\today}

\title{Non-local Correlation Effects in Fermionic Many-Body Systems: Overcoming the Non-causality Problem - Supplemental Material}

\maketitle

\section{The generalized cavity equations for nonlocal interactions}
The generalized DMFT equations derived in the main article considered nonlocal 
correlations in form of a nonlocal self-energy $\Sigma(k,i\omega)$. We can 
extend the derivation to nonlocal screenings in terms of a nonlocal polarization 
$P(q,i\nu)$ in the presence of nonlocal interactions. The general lattice action 
for a Hubbard model with nonlocal interactions is given by
\begin{align}
S 
&= \int_0^{\beta} \mathrm{d}\tau \sum_{i\sigma} c^*_{i\sigma}(\tau)\left[ \frac{\partial}{\partial \tau} - \mu \right] c_{i\sigma}(\tau) 
    -\sum_{ij,\sigma} \int_0^{\beta} \mathrm{d}\tau \ t_{ij} c^*_{i\sigma}(\tau)  c_{j\sigma}(\tau)  \nonumber \\
&\hspace{1cm} + \sum_{i} \int_0^{\beta} \mathrm{d}\tau \ \tilde{n}_{i\uparrow}(\tau) v_{ii} \tilde{n}_{i\downarrow}(\tau)
+ \sum_{i<j} \int_0^{\beta} \mathrm{d}\tau \ \tilde{n}_{i}(\tau) v_{ij} \tilde{n}_{j}(\tau),
\end{align}
where $c_{i\sigma}(\tau),c^*_{i\sigma}(\tau)$ are anticommuting Grassman 
variables, $\tilde{n}_{i\sigma}(\tau) = n_{i\sigma}(\tau) - 
\braket{n_{i\sigma}}$, $n_{i\sigma}(\tau) = 
c^*_{i\sigma}(\tau)c_{i\sigma}(\tau)$ and $\tilde{n}_{i}(\tau) = \sum_{\sigma} 
\tilde{n}_{i\sigma}(\tau)$. The hoppings $t_{ij}$ and interactions $v_{ij}$ are 
assumed to translationally invariant, e.g. $v_{ii}=v_{jj} \ \forall i,j$.

As discussed in the main text we split the action $S$ into three contributions
 $S  = S_0 + \Delta S + S^{(0)}$,
where $S_0$ is the term that contains only the local quantities on site $i=0$
\begin{align}
S_0
&= \int_0^{\beta} \mathrm{d}\tau \sum_{\sigma} c^*_{0\sigma}(\tau)\left[ \frac{\partial}{\partial \tau} - \mu \right] c_{0\sigma}(\tau) 
   + \sum_{\sigma} \int_0^{\beta} \mathrm{d}\tau \ t_{00} c^*_{0\sigma}(\tau)  c_{0\sigma}(\tau)  
 +  \int_0^{\beta} \mathrm{d}\tau \ \tilde{n}_{0\uparrow}(\tau) v_{00} \tilde{n}_{0\downarrow}(\tau).
\end{align}
The term $\Delta S$ contains all contributions that couples the site $i=0$ to all other sites 
\begin{align}
\Delta S 
&=  -\sum_{j\neq 0,\sigma} \int_0^{\beta} \mathrm{d}\tau \ \left( t_{0j} c^*_{0\sigma}(\tau)  c_{j\sigma}(\tau) + t_{j0} c^*_{j\sigma}(\tau)  c_{0\sigma}(\tau) \right) 
+ \sum_{j\neq 0 } \int_0^{\beta} \mathrm{d}\tau \ \tilde{n}_{0}(\tau) v_{0j} \tilde{n}_{j}(\tau).
\end{align}
And finally, $S^{(0)}$ contains all the contributions of the lattice with
site $i=0$ and its bonds connecting it being removed
\begin{align}
S^{(0)} 
&= \int_0^{\beta} \mathrm{d}\tau \sum_{i\neq 0,\sigma} c^*_{i\sigma}(\tau)\left[ \frac{\partial}{\partial \tau} - \mu \right] c_{i\sigma}(\tau) 
    -\sum_{ij\neq 0 ,\sigma} \int_0^{\beta} \mathrm{d}\tau \ t_{ij} c^*_{i\sigma}(\tau)  c_{j\sigma}(\tau)  \nonumber \\
&\hspace{1cm} + \sum_{i\neq 0 } \int_0^{\beta} \mathrm{d}\tau \ \tilde{n}_{i\uparrow}(\tau) v_{ii} \tilde{n}_{i\downarrow}(\tau)
+ \sum_{\substack{i<j \\ ij\neq 0  }} \int_0^{\beta} \mathrm{d}\tau \ \tilde{n}_{i}(\tau) v_{ij} \tilde{n}_{j}(\tau).
\end{align}
Now integrating out all the degrees of freedom except the ones on site $i=0$ one obtains 
an effective action of the form
\begin{align}
 S_{eff} &= S_0
 - \sum_{n=1}^{\infty} \frac{(-1)^n}{n!} \braket{ (\Delta S)^n }^{(0)},
\end{align}
where $\braket{  }^{(0)} = \frac{1}{\mathcal{Z}^{(0)}} \int \prod_i 
\mathrm{d}c^*_{i\sigma} \mathrm{d}c_{i\sigma} \mathrm{e}^{-S^{(0)}}$ indicates a 
trace over the system with the site $i=0$ removed with a corresponding partition 
function $\mathcal{Z}^{(0)}$. In Ref.\cite{Si1996} it has been shown that with a 
rescaling of $t_{ij} \rightarrow t_{ij}/\sqrt{2d}$ and $v_{ij} \rightarrow 
v_{ij}v_{00}/\sqrt{2d}$, where $d$ is the dimension of the lattice, that in the 
effective action all terms $n>2$ vanish in the limit of $d\rightarrow \infty$. 
The first order terms vanishes because the hopping term contains only 
expectation values of one Grassman variable, and the first order interaction 
term is proportional to $\braket{ \tilde{n}_{i\sigma} }^{(0)} = \braket{ 
n_{i\sigma} }^{(0)} - \braket{n_{i\sigma}}$, which is of order $1/d$. 
Additionally, different orders $n$ have vanishing interference terms between the 
fermionic $t_{ij}$ and density $v_{ij}$ terms, thus there will be two separate 
series of terms for the $t_{ij}$ and $v_{ij}$ contributions. Thus, taking into 
account only the zeroth and the second order contribution we obtain an 
approximate form of the effective action which is becomes exact in the 
$d\rightarrow \infty$ limit and evaluates to
\begin{align}
S_{eff} 
&= S_0 - \frac{1}{2} \braket{ (\Delta S)^2 }^{(0)} \nonumber \\
&= \int_0^{\beta} \mathrm{d}\tau_1 \int_0^{\beta} \mathrm{d}\tau_2 \sum_{\sigma} c^*_{0\sigma}(\tau_1)\left(
                \delta(\tau_1-\tau_2)\left[\frac{\partial}{\partial \tau} - \mu+t_{00} \right]
                + \sum_{ij\neq 0} t_{0i} t_{j0} G_{ij\sigma}^{(0)}(\tau_1-\tau_2)   \right) c_{0\sigma}(\tau)  \nonumber \\
&\hspace{1cm}    + \int_0^{\beta} \mathrm{d}\tau_1 \int_0^{\beta} \mathrm{d}\tau_2 \ \tilde{n}_{0}(\tau_1) 
       \left(\delta(\tau_1-\tau_2) v_{00} - \sum_{ij\neq 0} v_{0i} v_{j0} \chi^{(0)}_{ij}(\tau_1-\tau_2) \right)   \tilde{n}_{0}(\tau_2)  .
\end{align}
where we have introduced the cavity quantities
\begin{align}
 G_{ij\sigma}^{(0)}(\tau_1-\tau_2) &= -\braket{T_{\tau} c_{i\sigma}(\tau_1) c^*_{j\sigma}(\tau_2) }^{(0)} \\
 \chi_{ij}^{(0)}(\tau_1-\tau_2) &= \braket{ \tilde{n}_{i}(\tau_1)   \tilde{n}_{j}(\tau_2) }^{(0)} .
\end{align}
This expression then takes the form of an impurity action 
\begin{align}
S_{eff} 
&= -\int_0^{\beta} \mathrm{d}\tau_1 \int_0^{\beta} \mathrm{d}\tau_2 \sum_{\sigma} 
          c^*_{0\sigma}(\tau_1)\ \mathscr{G}^{-1}_{\sigma}(\tau_1-\tau_2) c_{0\sigma}(\tau)  
    + \int_0^{\beta} \mathrm{d}\tau_1 \int_0^{\beta} \mathrm{d}\tau_2 \ \tilde{n}_{0}(\tau_1) 
       U(\tau_1-\tau_2)  \tilde{n}_{0}(\tau_2)  ,
\end{align}
with the effective Weiss field and interaction given by
\begin{align}
 \mathscr{G}^{-1}_{\sigma}(\tau_1-\tau_2) 
 &= -\delta(\tau_1-\tau_2)\left[\frac{\partial}{\partial \tau} - \mu +t_{00} \right] - \sum_{ij\neq 0} t_{0i} t_{j0} G_{ij\sigma}^{(0)}(\tau_1-\tau_2) \\
 U(\tau_1-\tau_2) 
 &= \delta(\tau_1-\tau_2) v_{00} - \sum_{ij\neq 0} v_{0i} v_{j0} \chi^{(0)}_{ij}(\tau_1-\tau_2).
\end{align}
Or in the frequency domain
\begin{align}
 \mathscr{G}^{-1}_{\sigma}(i\omega) 
 &= i\omega + \mu - t_{00} - \sum_{ij\neq 0} t_{0i} t_{j0} G_{ij\sigma}^{(0)}(i\omega) \\
 U(i\nu) 
 &= v_{00} - \sum_{ij\neq 0} v_{0i} v_{j0} \chi^{(0)}_{ij}(i\nu). \label{eq:eff_interaction_cavity}
\end{align}
The connection to the original lattice including site $i=0$ can be made by using the  equations
\begin{align}
 G^{(0)}_{ij} &= G_{ij} - G_{i0}G^{-1}_{00}G_{0j} \\ 
 \chi^{(0)}_{ij} &= \chi_{ij} - \chi_{i0}\chi^{-1}_{00}\chi_{0j} ,
\end{align}
suppressing the spin index for readability. As discussed in the main text, one 
then arrives at the following expression for the Weiss field
\begin{align}
\mathscr{G}^{-1} (i\omega)
 &= i\omega + \mu - \braket{\epsilon} - \Big( \braket{ \epsilon G \epsilon } - \braket{\epsilon G} \braket{G}^{-1}  \braket{ G \epsilon }  \Big) \nonumber  \\
 &= \braket{G}^{-1} + \braket{\Sigma} 
  - \Big( \braket{\Sigma G \Sigma} - \braket{\Sigma G} \braket{G}^{-1} \braket{G\Sigma} 
     + 2\braket{\Sigma} - \braket{\Sigma G} \braket{G}^{-1} - \braket{G}^{-1}\braket{G\Sigma}\Big) .
\end{align}
with $\braket{}$ indicating a local projection.
 
When comparing the effective interaction in Eq.\eqref{eq:eff_interaction_cavity} 
to the fermionic bath, we see that the bare interaction $v$ formally plays the 
same role as the noninteracting hopping amplitude $t$, and the susceptibility 
$\chi$ takes the role of the single-particle Green's function $G$. Therefore, 
following the same derivation as for the single-particle Green's function the 
analogous result for the effective interaction is 
\begin{align}
 U(i\nu) 
 = \braket{v} - \left( \braket{ v \chi v } - \braket{v \chi} \braket{\chi}^{-1} \braket{\chi v} \right) . 
 \label{eq:eff_dyson_for_interaction}
\end{align}
This equation  can be rewritten in terms of the screened interaction $W$ and polarization $P$,
by making use of the following identities
\begin{align}
 \chi &= - P -PWP \\
 v \chi &= -PW \\
 \chi v &= -WP \\
 v\chi v &= v - W,
\end{align}
which leads to the following expression
\begin{align}
 U(i\nu) 
&= [\braket{W}^{-1}+\braket{P}]^{-1} - \Big( \braket{PW} \big[ \braket{P}+\braket{PWP} \big]^{-1} \braket{WP}
      - \braket{P}\braket{W} \big[\braket{P}+\braket{P}\braket{W}\braket{P} \big]^{-1} \braket{W} \braket{P} \Big) .
      \label{eq:eff_int_general}
\end{align}
We see that we obtain the usual local Dyson-like equation in the first term, but 
with an additional correction term. The correction term vanishes in case of a 
purely local polarization $P(q,i\nu)=P(i\nu)$, since the local projection 
separates $\braket{PW}= \braket{P} \braket{W}$ and the correction terms cancel 
each other, and we obtain the local Dyson-like equation
\begin{align}
 U(i\nu) &=  \left[ \braket{W}^{-1} + \braket{P} \right]^{-1}.
 \label{eq:eff_int_dyson}
\end{align}

\begin{figure*}[t]
\includegraphics[width=1\textwidth]{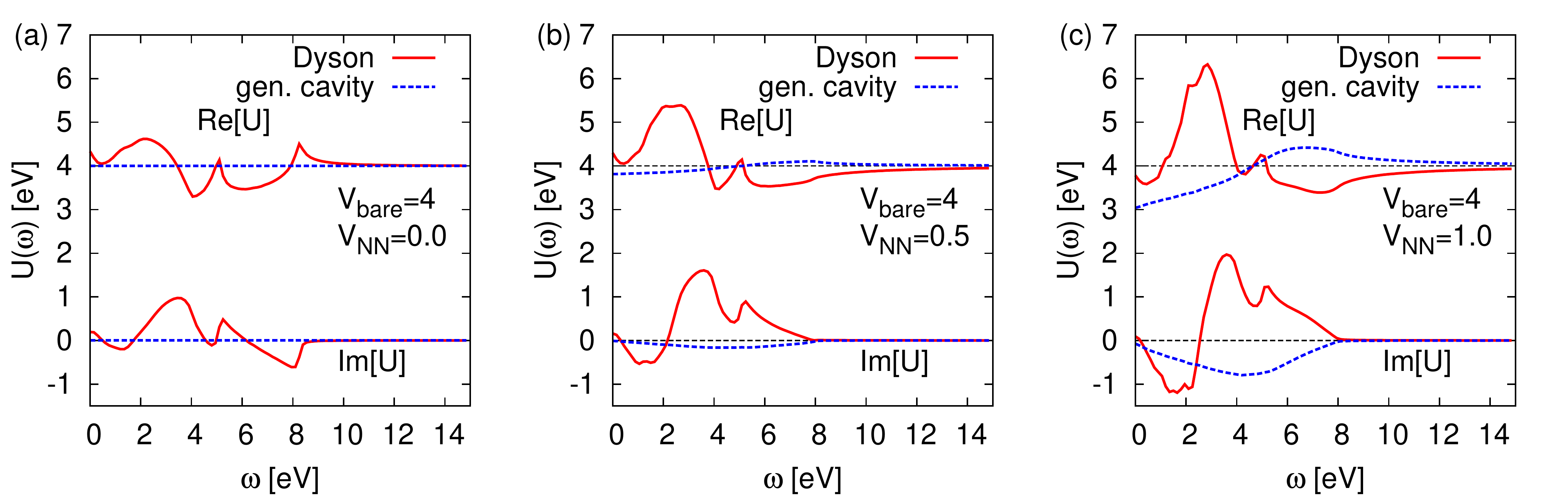} 
\caption{The effective impurity interaction $U(\omega)$ on real frequencies $\omega$ for a two-dimensional single-orbital Hubbard model at half-filling for
hopping $t=1$, bare on-site interaction $V_{bare}=4t$ for different values of the nearest-neighbor interaction 
$V_{NN}$. The polarization was obtained within the RPA approximation $P=G_0G_0$. The interaction generated by
the Dyson-like equation Eq.\eqref{eq:eff_int_dyson} shows noncausal behavior, indicated by $\mathrm{Im}[U]>0$, and unphysical screening effects
for $V_{NN}=0$. The generalized cavity equation Eq.\eqref{eq:eff_int_general} generates a causal interaction and obtains 
the correct limit of the bare interaction for the case of $V_{NN}=0$.
}
\label{fig:eff_int}
\end{figure*}
Similar to the effective impurity bath Green's function $\mathscr{G}$, the 
effective impurity interaction $U$ when generated from the Dyson-like equation 
Eq.\eqref{eq:eff_int_dyson} will show in general noncausal features in the 
presence of nonlocal polarization. Again, the additional term in 
Eq.\eqref{eq:eff_int_general} ensures causality as $\chi^{(0)}_{ij}$ is causal 
and thus implies causality of $U$ if no further approximations are introduced. 
As an example we show the resulting effective interaction $U(\omega)$ for a 
two-dimensional single-orbital Hubbard model at half filling with hopping $t=1$ and 
on-site bare interaction $V_{bare}=4t$ for different values of a 
nearest-neighbor interaction $V_{NN}$ in Fig.\ref{fig:eff_int}. The 
polarization to screen the bare interaction has been generated using the RPA 
approximation $P=G_0G_0$. Comparing the results of Eq.\eqref{eq:eff_int_general} 
and Eq.\eqref{eq:eff_int_dyson} we observe significant noncausal features in the 
effective interaction when the Dyson equation is used, indicated by positive 
values of the imaginary part of $U$. Here, the noncausal weight is in fact 
larger than the causal one. Additionally we observe an unphysical screening 
effect for a purely on-site interaction (Fig.\ref{fig:eff_int}(a)) from the 
Dyson-like equation, which might explain similar reports in self-consistent dual 
boson calculations\cite{Stepanov2016scDB}. On the other hand, 
Eq.\eqref{eq:eff_int_general} generates a causal effective interaction, 
recovering the bare interaction for $V_{NN}=0$, and enhanced screening for 
increasing nearest-neighbor interaction.

\section{Nonanalyticity of the impurity model}
\begin{figure*}[t]
\includegraphics[width=1\textwidth]{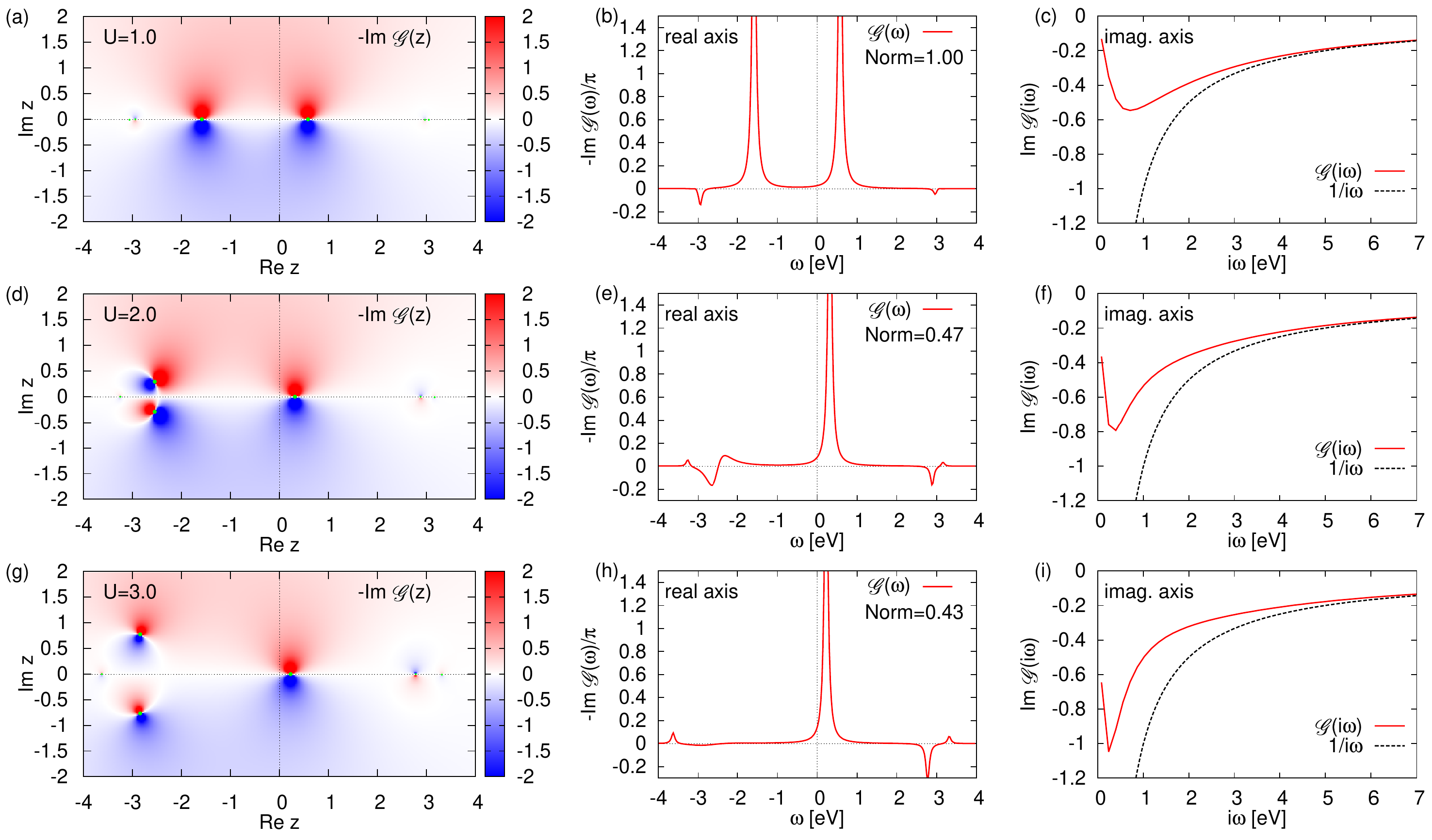} 
\caption{The effective single-site impurity bath Green's function $\mathscr{G}$ for the two-site dimer, generated from the exact solution
using the impurity Dyson equation for the interaction (a)-(c) $U=1t$, (d)-(f) $U=2t$ and (g)-(i) $U=3t$.
The spectrum shown on the real axis corresponds to the limit in the upper complex plane $\omega + i0^+$.
The poles of the Green's function in the complex plane are located between the red/blue extremal values, marked with a green dot.
For small interactions the bath shows negative spectral weight but is normalized to one electron, since all poles of the Green's
function are located on the real axis.
For larger interaction values the poles of the Green's function move away from the real axis into the complex plane,
reducing the normalization. Though, the Matsubara Green's function still shows a $1/i\omega$ behavior, indicating
a violation of the Hilbert transform.
}
\label{fig:nonanalytic_bath_map}
\end{figure*}
The problem of the emergence of a nonanalytic impurity solution in the 
upper/lower complex plane when the bath is generated from the Dyson equation is 
not related to the type of approximations involved, but occurs in general even 
for the exact solution. In order to see this and the resulting nonanalyticity 
(i.e., poles entering the complex plane), we show the single-site impurity bath 
Green's function $\mathscr{G}$ for a dimer as discussed in the main text in the 
whole complex plane, generated from the exact self-energy in 
Fig.\ref{fig:nonanalytic_bath_map}. For all interaction values considered 
noncausal spectral weight, represented by negative values on the real axis, is 
present. For small interactions all poles of the Green's function are located on 
the real axis, but some of them have negative weight, as can be seen in 
Fig.\ref{fig:nonanalytic_bath_map} (a) and (b) at $\omega=\pm 3$~eV. This 
negative spectral weight corresponds to negative hybridization amplitudes 
$|V|^2$, i.e. a purely imaginary and non-hermitian matrix element coupling the 
impurity to the bath.

As the interaction increases, two poles on the real axis merge and move into the 
complex plane, corresponding to two complex Eigenvalues $E_{\pm} = E_r \pm 
iE_i$. As a result the spectrum is not only noncausal as before, but also no 
longer normalized on the real axis. At the same time no features indicating such 
a behavior are visible in the Matsubara Green's function on the imaginary axis, 
which always shows a $1/i\omega$ decay for all interaction values. This 
represents a breakdown of the Hilbert transform
\begin{align}
 G(i\omega_n) &= \int \frac{A(\omega)}{i\omega_n - \omega } \mathrm{d}\omega,
\end{align}
which links the real frequency spectrum $A(\omega)$ to the Matsubara Green's function $G(i\omega_n)$. 
For large frequencies one obtains the relation between the normalization of the spectrum
and the high-frequency behavior on the Matsubara axis
\begin{align}
 G(i\omega_n) &\approx \frac{1}{i\omega_n} \int A(\omega) \mathrm{d}\omega, \hspace*{0.5cm} \mbox{ for large }\omega_n,
\end{align}
which is clearly violated for larger interactions in 
Fig.\ref{fig:nonanalytic_bath_map} (f) and (i), as the tail of 
$\mathscr{G}(i\omega)$ indicates a normalization on the real axis equal to one. 
Such behavior is expected as the Hilbert transform only holds as long as the 
Green's function is analytic in the upper/lower complex plane, which is not the 
case when poles separate from the real axis into the complex plane. This is 
problematic as the real-frequency spectrum cannot be reconstructed by analytic 
continuation from the Matsubara Green's function by methods commonly used like 
the maximum entropy method~\cite{maxent}, which relies on the Hilbert transform. 
Though, the Pad\'e approximation\cite{Pade1892} does not face this restriction 
and can still be applied, but is limited in practice due to a high sensitivity 
on numerical noise. We usually observed that the impurity solution, i.e. the 
impurity Green's function and self-energy exhibited even stronger nonanalytic 
behavior, with more poles leaving the real axis. As discussed in the main text, 
for $U=3t$ all poles for the Selfenergy had left the real axis.

\begin{figure*}[t]
\includegraphics[width=1\textwidth]{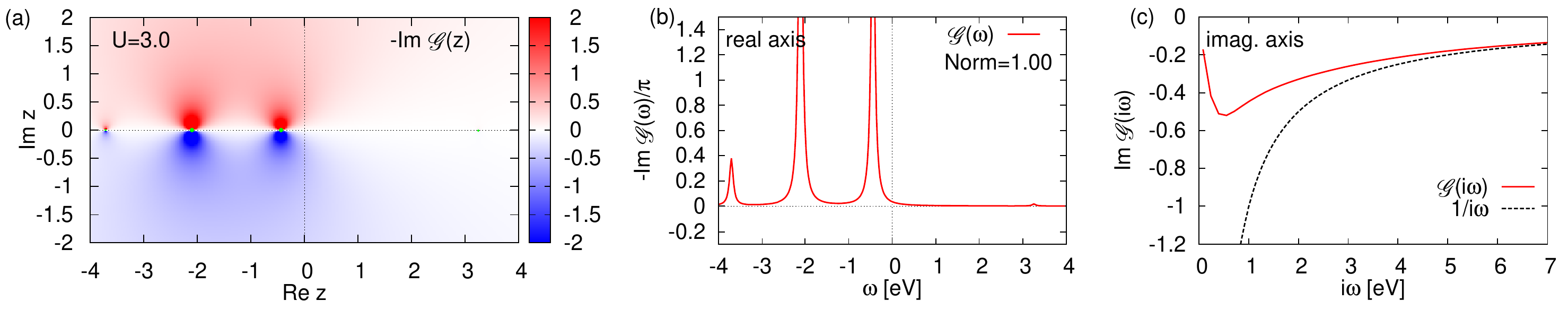} 
\caption{The effective single-site impurity bath Green's function $\mathscr{G}$ for the two-site dimer
as in Fig.\ref{fig:nonanalytic_bath_map}, but generated from the generalized cavity DMFT equation Eq.(5) in the main text
for $U=3t$. No noncausal spectral weight is present and all poles of the Green's function are located on the real axis.
}
\label{fig:analytic_bath_map}
\end{figure*}

As described in the main text, the generalized cavity DMFT equations are causal. 
This is confirmed in Fig.\ref{fig:analytic_bath_map}, which shows the effective 
bath Green's function for $U=3t$. The spectrum is positive, normalized to one 
electron and the relation between the real and imaginary axis via the Hilbert 
transform still holds as the function is analytic in the whole upper 
complex/lower plane. This applies to all other derived quantities like the 
hybridization, impurity Green's function or self-energy.

%